\documentclass[11pt,twocolumn]{article}

\pdfoutput=1

\usepackage[utf8]{inputenc}
\usepackage{tcolorbox,multicol}
\usepackage{mathtools,amssymb}
\usepackage{listings,booktabs}
\usepackage{tabularx}
\usepackage{soul}
\usepackage{cleveref}
\usepackage[font={sf,normalsize},labelfont={sf}]{caption}
\usepackage{microtype}
\usepackage{cuted}
\usepackage[numbers]{natbib}
\usepackage[switch]{lineno}
\usepackage{titlesec}

\titlespacing*{\section}{0pt}{*1}{0pt}
\titlespacing*{\subsection}{0pt}{*1}{3pt}

\usepackage[margin=0.65in]{geometry}

\usepackage{setspace}
\usepackage{lmodern}

\title{
Gradient Diffusion:
Sensitivity-Matrix Co-Simulation Enables
Activity Adaptation and Learnable Plasticity in Neural Simulators
}
\author{
Lennart P. L. Landsmeer${}^{1,2}$,
Mario Negrello${}^2$,
Said Hamdioui${}^1$,
Christos Strydis${}^{2,1}$
\\
\\
{
\small
${}^1$
Department of Quantum and Computer Engineering, 
Delft University of Technology,
Delft, The Netherlands
}
\\
{
\small
${}^2$
NeuroComputingLab,
Department of Neuroscience,
Erasmus Medical Center,
Rotterdam, The Netherlands
}
}
\date{\small December 2025}

\begin{document}

\maketitle

\newcommand{\calcjac}[2]{\frac{\partial #1}{\partial #2}{}}
\newcommand{\calcjacovercm}[2]{\frac{\partial #1}{C_m\partial #2}{}}
\newcommand{\varjac}[2]{\left[\hspace{-1pt}\frac{\partial #1}{\partial #2}{}\hspace{-1pt} \right]}
\newcommand{\dt}{\delta t}
\newcommand{\dSdx}{\left(\frac{\partial S}{\partial x}\right)}

\begin{abstract}
{
Computational neuroscience relies on large-scale dynamical-systems models of neurons, with a vast amount of offline, pre-simulation, tuned parameters, with models often tied to their brain simulators.
These fixed parameters lead to stiff models, that show unnatural behaviour when introduced to new environments, or when combined into larger networks.
In contrast to offline tuning, in biology,  cells continuously adapt via homeostatic plasticity to stay in desired dynamical regimes.
In this work, we aim to introduce such online tuning of cellular parameters into brain simulation. We show that the sensitivity equation of a biorealistic neural models has the same shape as a general neuron model, and can be simulated within existing brain simulators.
Via co-simulation with the sensitivity equation, we enable both offline, and online tuning of activity of arbitrary biophysically realistic brain models.
Furthermore, we show that this opens the possibility to study the biological mechanisms underlying homeostatic plasticity, via both meta-learning plasticity mechanism as well as treating online tuning as a black-box plasticity mechanism.
Through the generality of our methods, we hope that more computational science fields can capitalize on the similarity between the simulated model and its gradient system.
}
\end{abstract}

\begin{figure*}[t!]
    \centering
    \includegraphics[width=1.0\textwidth]{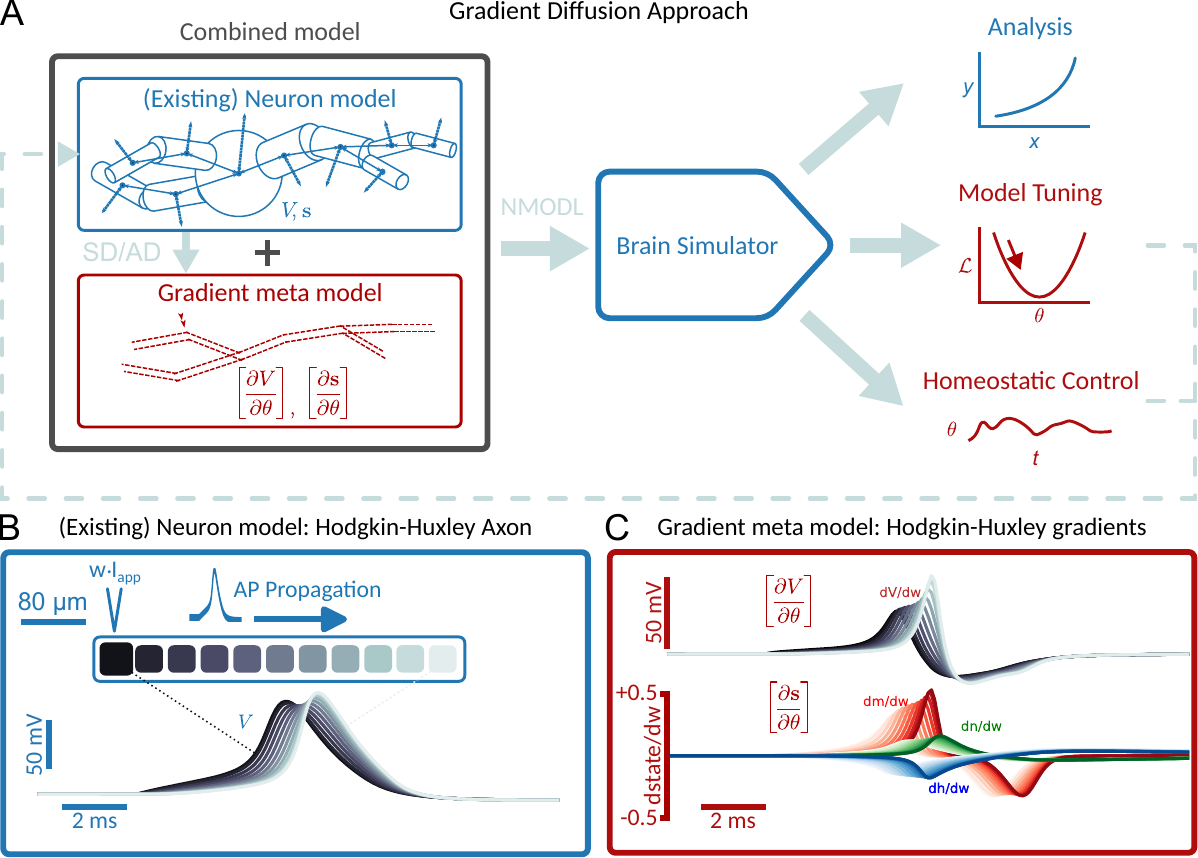}
    \caption{
        Gradient diffusion via gradient meta-models enhance existing brain simulators with gradient calculation for creating flexible and composable brain models.
        \textbf{A}
        Gradient-diffusion approach:
        A single neuron model is enhanced with a gradient (meta-) model, by symbolic or automatic differentiation.
        This renders an extended model that can be conveniently co-simulated by existing brain simulators.
        Resulting gradients can be used for automated tuning, homeostatic control, or be useful on their own.
        A standard biophysically detailed neuron model is commonly modeled as a discretized, multicompartmental, cable cell. Voltage diffusion happens across compartments (connected arrows), and via ion-channel/leak-based axial currents (dotted arrows). Single compartment state variables including state derivatives and transmembrane currents, are commonly written in NMODL files.
        \textbf{B}
        Hodgkin-Huxley (HH), linear axon model, with a pulse current applied on one end to create a propagating action potential (AP).
        Each compartment is represented in a different shade of gray, indicating AP propagation.
        \textbf{C}
        Gradient model of the HH model, showing the resulting gradients for all state variables over time.
    } \label{fig:cablecelldef}
\end{figure*}

\section*{Introduction}
In computational neuroscience,
explaining large-scale brain activity relies on models that link circuit dynamics directly to the biophysical mechanisms that generate them.
These conductance-based neuron models require extensive tuning of cellular model-parameters, to generate behaviour matching experimentally observed biology.
Still, such models remain brittle, overfitted, and static, and
when combined into larger networks, need to be retuned at the network-level, which does not scale~\cite{almog2016realistic}.
Biology solves these problems via the online adaptation of neural activity to environmental change, also known as homeostatic plasticity. 
Inspired by this, we suggest that
brain models require a similar scalable, online, automatic tuning mechanism, both as a method 
to overcome this brittle neuron problem, as well as a method to study the slow behaviour of neurons.

Existing offline
parameter optimization methods in current simulators rely almost exclusively on gradient-free methods such as evolutionary search or random sampling. These approaches scale poorly in high dimensions and cannot support online adaptation during simulation, where gradients are required for efficient homeostatic control. Although recent automatic-differentiation simulators have emerged, they remain incompatible with established frameworks and existing models, and, still do not support online tuning of neural models. Online control mechanisms have been manually developed in certain cases, but not as a general solution in brain simulation.

Here, we introduce a general gradient-based tuning framework that brings online adaptability to existing biophysical brain models. Our method computes parameter gradients for any unmodified model–simulator combination, enabling both offline optimization and biologically inspired online adaptation. This capability allows existing simulators to implement self-tuning and metaplasticity mechanisms, endowing detailed conductance models with scalable learning dynamics and opening a promising path toward adaptive and composable brain simulations.

\begin{figure*}[t!]
    \centering
    \includegraphics[width=1.0\textwidth]{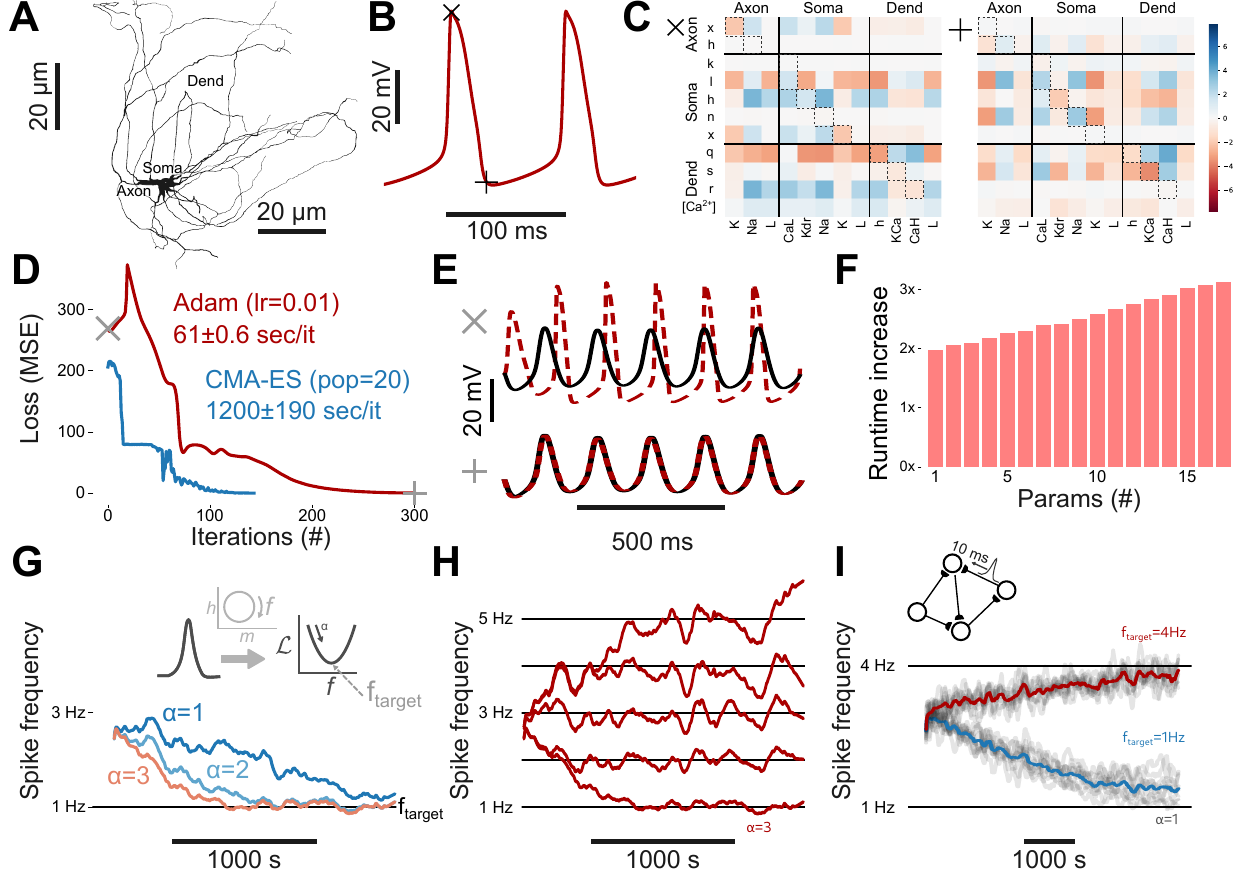}
    \caption{
        Gradient-diffusion enables offline and online tuning of biophysical realistic neuron models.
        \textbf{A:} The morphology of the realistic multicompartment neuronal model used. This model was chosen for its complex dynamics with subthreshold oscillations, cross-membrane calcium diffusion, calcium plateaus and differential ion-channel distributions.
        \textbf{B:} Subthreshold Oscillation (STO) with specific timepoints indicated ($+$ and $\times$) and shown in C 
        \textbf{C:} Gradients for all state variables ($\varjac{\mathbf s}{\theta}$) at specific timepoints from B.
        \textbf{D:} Loss during optimization using gradient descent (Adam, red) and the traditional evolutionary approach (CMA-ES, blue)
        \textbf{E:} Simulated (red) vs target (black) somatic membrane potential before (top) and after (bottom) training
        \textbf{F:}
        Increase in computational time relative to non-gradient simulation, when co-simulating the gradient model.
        \textbf{G:} Homeostatic control of oscillation frequency via online learning. Frequency estimation is derived from $(m,h)$-state trajectories (see Methods). Traces show the effect of learning rate ($\alpha$, in $\cdot 10^{-6}$) on homeostatic control stability of frequency, with $\alpha>3\cdot 10^{-6}$ being unstable.
        The simulation is 2000 seconds, or 400 million time-steps.
        \textbf{H:} Homeostatic control with different target frequencies, same as G.
        \textbf{I:} 
        Online network tuning via homeostatic control. A network of 10 Hodgkin-Huxley neurons is simulated, where each neuron is receiving spike input from 3 other random neurons. The system tunes to the desired frequency even in the absence of spike-gradients.
    } \label{fig:sto}
\end{figure*}

\section*{Results}

\begin{figure*}[t]
    \centering
    \includegraphics[width=1.0\textwidth]{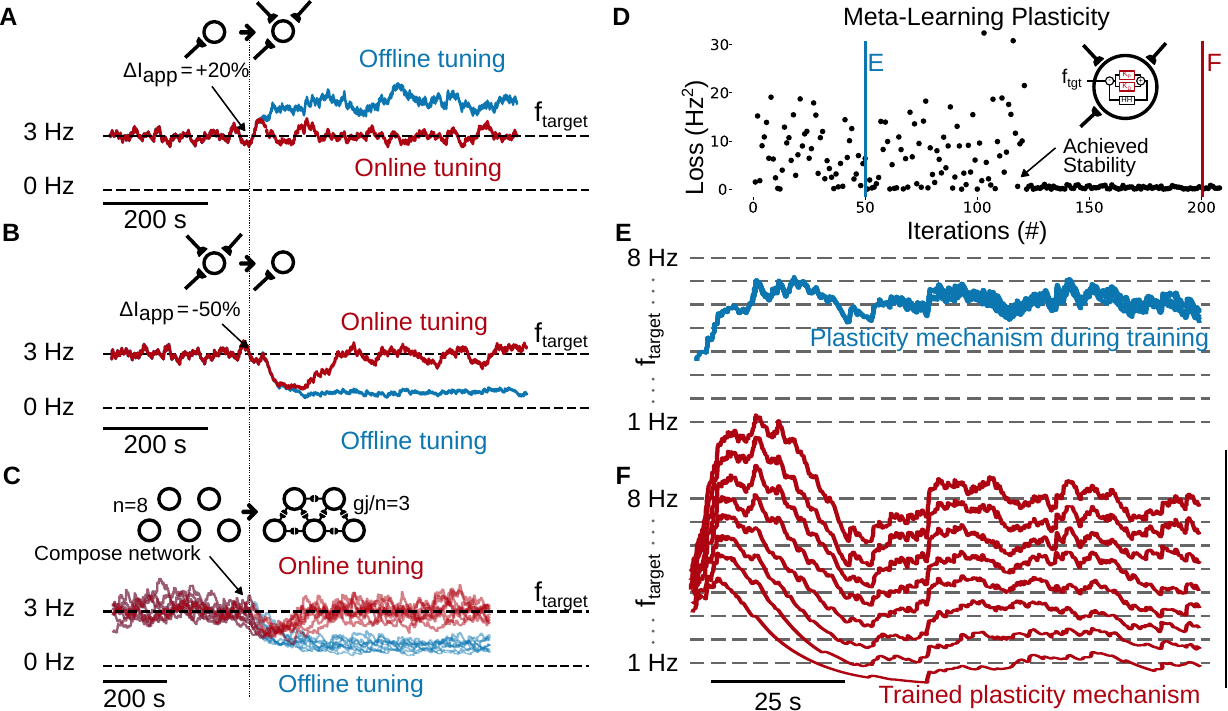}
    \caption{
    Online tuning via gradient diffusion makes brain simulations more adaptable and composable (A-C), and gradient diffusion allows meta-learning plasticity mechanisms inside cells (D-F).
    \textbf{A, B}: Adoptation to increase (A) or decrease (B) in stimuli. While offline tuned cells can maintain their tuned behaviour as long as the environment does not change, subte variations lead to non-desired behaviour.
    Automated online tuning, using gradients (this method), keep cells in their desired target state, similar to biological homeostasis.
    \textbf{C}: Composition of neural mdoels after connecting neurons in a network via gap junctions is known to alter the neurons behaviour, making it tedious to find the right value for the channel conductance values in a network. Via online tuning, we can find the right values for the desired behaviour in a connected network, making neural models more composable.
    \textbf{D}: Meta-learning a homeostatic plasticity mechanism as a PD controller over 200 iterations. Blue and red vertical bars correspond to E and F.
    \textbf{E}: Initial homeostatic plasticity mechanism.
    \textbf{F}: Homeostatic plasticity mechanism after training.
    } \label{fig:improve}
\end{figure*}

\begin{figure*}[t]
    \centering
    \includegraphics[width=1.0\textwidth]{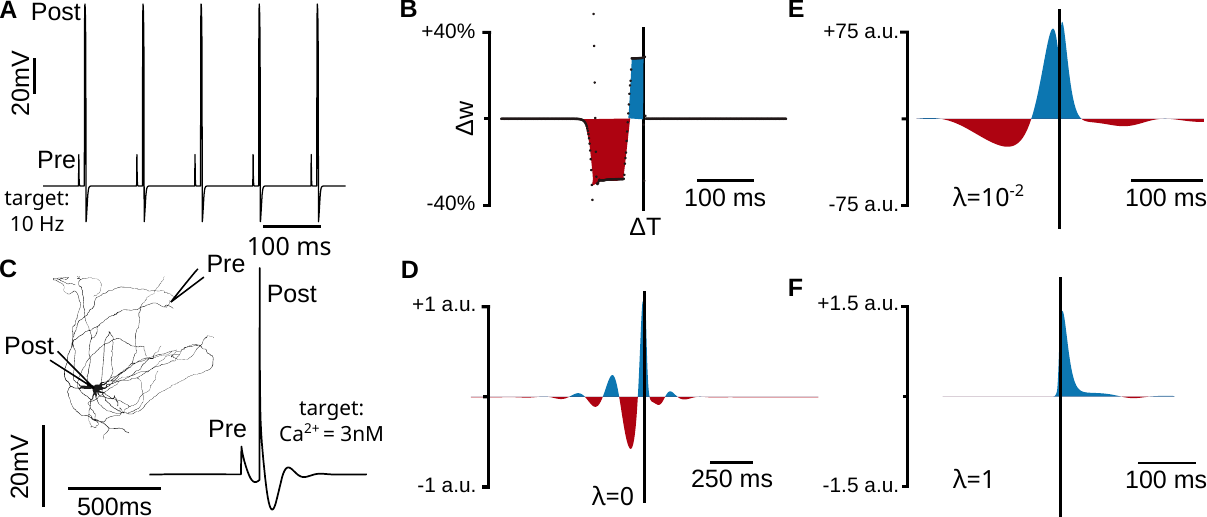}
    \caption{
    Spike-timing dependent plasticity as a by-product of automatic homeostatic control
    \textbf{A}: Induced stimulation protocol in the HH model. The control process aims to bring the cell to 10 Hz spiking.
    \textbf{B}: Resulting weight changes after 10s of 10Hz stdp protocol, for various pre/post delays. Black dots are included to show all values.
    \textbf{C}: Inferior olive neuron, with pre- and post stimulation sites. The system is tasked with optimizing the Ca$^{2+}$-concentration towards 3nM.
    \textbf{D,E,F}: Resulting STDP response, for varying values of the forgetfulness parameter $\lambda$. Note how the resonating subthreshold response to the presynaptic spike, results in a oscillatory STDP pattern for low values of $\lambda$.
    } \label{fig:stdp}
\end{figure*}

Let us assume a morphologically detailed, conductance based, biophysically realistic model of a single neuron (\cref{fig:cablecelldef}A-B). Within the brain-simulation software environment a modeler essentially defines two sets of functions: the ion-channel current contributions $i$ to the membrane potential $V$,
and the time-derivative of the internal state variables $\mathbf s$.
These functions are defined for each position on the spatial neuron and depend on a set of parameters $\theta$:

\begin{align}
        i\left(V, \mathbf s, \theta \right) \label{eq:idef}\\
    \frac{d}{dt} {\mathbf s} \left(V, \mathbf s, \theta \right)
            \label{eq:model}
\end{align}

\noindent On the basis of this description, our goal is to find the gradients of the voltage $V$ and state $\mathbf s$ with respect to the parameters $\theta$ at any point in time and for all compartments. To that end, we define a \emph{gradient model} to be simulated alongside the neural model in \cref{eq:idef} and \cref{eq:model}. This gradient model consists of the voltage gradients $\varjac{V}{\theta}$ as diffusive ion-species and the state gradients $\varjac{\mathbf s}{\theta}$ as internal state variables (\cref{fig:cablecelldef}A-C). In our notation, brackets ($\left[\cdot\right]$) denote new state variables, in fact, components of the sensitivity matrix, introduced for the simulation of the gradient model. The gradient model calculates the voltage and state gradients via direct integration of the sensitivity equation (see Methods) by supplying the time derivatives of
$\varjac{V}{\theta}$ and $\varjac{\mathbf s}{\theta}$ as mechanisms to the brain simulator:

\begin{align}
    \frac{d}{dt}
    {\varjac{V}{\theta}}
    \left(
    V,
    \mathbf s,
    \theta,
    \varjac{V}{\theta},
    \varjac{\mathbf s}{\theta}
    \right)
    \\
    \frac{d}{dt}
    {\varjac{\mathbf s}{\theta}}
    \left(
    V,
    \mathbf s,
    \theta,
    \varjac{V}{\theta},
    \varjac{\mathbf s}{\theta}
    \right) \label{eq:gradmodel}
\end{align}

\noindent These time derivatives can be obtained by either symbolic or automatic differentiation. Specifically, the gradient model can be fully expressed either as existing NMODL interfaces (\cref{fig:nmodl}) or via the Arbor-PyCat interface developed in Methods, respectively.

\subsection*{Applications}

We now demonstrate how the co-simulation of gradients in brain models, in the form of the sensitivity matrix,
improves over existing techniques in offline tuning and allows for novel online \emph{automatic homeostatic control} and \emph{meta-learning of plasticity}, leading to \emph{adaptable and composable brain models}, and finally using automatic homeostatic control as a \emph{stand-in for unknown learning processes} in the brain.

\paragraph{Gradient-Based Parameter Tuning}

Given a Loss function $\mathcal L \left(V, \mathbf{s}\right)$, we can define gradient descent by integration over the total simulation time $T$ as:

\begin{align}
    \theta
    \leftarrow
    \theta -
    \alpha
    \int_0^T
    \left(
    \frac{\partial \mathcal {L}}{\partial V}
    \varjac{V}{\theta}
    +
    \frac{\partial \mathcal {L}}{\partial \mathbf s}
    \varjac{\mathbf s}{\theta}
    \right)
    dt
    \label{eq:gupdate}
\end{align}

\noindent Here, the application of gradients in brain simulation is external tuning of parameters.
To show the applicability of the method, we developed the gradient-model of a 17-parameter Inferior Olivary Nucleus (ION) neuron model \cite{de2012climbing,negrello2019quasiperiodic} with experimentally obtained morphology 
 (\cref{fig:sto}A). These neurons can exhibit sustained subtreshold oscillations (\cref{fig:sto}B). The gradients show how the interplay between parameters and state gives rises to the observed behaviour (\cref{fig:sto}C). We optimized the aforementioned ION model using gradient-descent \cite{kingma2014adam,deepmind2020jax} to replicate a realistic sinusoidal, target voltage trace. For comparison, traditional evolutionary search (CMA-ES~\cite{hansen2019pycma}) was also included. Both optimization methods reach a solution, but CMA-ES relies on more simulations per iteration, leading to $\approx$10 times longer optimization times (\cref{fig:sto}D-E). Co-simulating a gradient system with the regular system for simulation increased simulation runtime by $\approx$2 times, after which, minor runtime increase per included parameter is observed \cref{fig:sto}F.

\paragraph{Automatic Homeostatic Control}

The principal advantage of the co-simulated gradient model over externally calculated gradients, is access to gradients directly within the neural model. These can be used for \emph{online} optimization of neurons; i.e., for emulating homeostatic control. This will allow us to tune whole networks of cells during runtime.

The goal of homeostatic control is to minimize an instantaneous error  $e\left(V, \mathbf s\right)$. For simplicity, we can assume that this error is a single element of $\mathbf s$, the integrated loss term $\mathbf s_0 = \int_0^t l(v, \mathbf s)$ such that $e(V, \mathbf s) = \mathbf s_0$.
Now, we can interpret gradient descent as an dynamical system with learning rate $\alpha$:

\begin{align}
    \frac{d}{dt} {\mathbf \theta} &=
        - \alpha \frac{\partial e\left(\mathbf s\right)}{\partial \mathbf s } \varjac{\mathbf s}{\theta}
        +\xi \Delta_x \theta
        \label{eq:paramsovertime}
\end{align}

\noindent The parameters are now diffusive ion species, where fast diffusion of gradient updates $\xi$ enables non-local tuning. A gradient-forgetting parameter $\lambda$ is not required to accommodate for the updated values of $\theta$ which are not yet reflected in $\varjac{\mathbf s}{\theta}$. We add this term to make sure the gradients do not explode for long running simulations. We denote these adjusted gradients with a prime ($'$):

\begin{align}
\frac{d}{dt} {\varjac{a}{\theta}}'
&=
-\lambda {\varjac{a}{\theta}}' +
\frac{d}{dt} {\varjac{a}{\theta}},\quad a\in\left\{V, \mathbf s\right\}\label{eq:forget}
\end{align}

We applied the homeostatic control as described above to the Hodgkin-Huxley system, to tune the frequency of single neurons. Frequency was derived from the state dynamics (\cref{fig:sto}G, top). Homeostatic control enables Hodgkin-Huxley neurons to tune, during simulation runtime, their frequency response in the presence of a varying noise input.
Given appropriate balancing of $\alpha$ and $\lambda$, this leads to a stable system (see \cref{fig:sto}G, Methods).
The same method can be applied to tune for a range of target frequencies (\cref{fig:sto}H).
In the network setting, homeostatic tuning also stably enables frequency-tuning, although at a lower learning rate (\cref{fig:sto}I).

\paragraph{Activity Adaptation for Adaptable and Composable Brain Models}
Current brain simulations are often touted as brittle `glass models', that show the desired behaviour solely in their regime of interest~\cite{almog2016realistic}. Introduction of brain models in new environments, new compositions of neurons, addition of new areas, or plasticity-driven changes in input often break these models.
Homeostasis makes neurons behave similarly across varying environments in biology, but currently not in simulations, as shown by this brittleness. Introducing automatic homeostasis to neurons, brain simulations become more adaptable and composable. For instance, when considering neurons that should have a spiking frequency of around 3 Hz, changes in connectivity bring the neurons behaviour outside its biological operation range. When automatic homeostasis is employed, the neuron can maintain its healthy frequencies, under addition, removal or composition of networks.
This leads to more scalable brain models, than can be combined and composed across vastly larger scales than currently possible.

\paragraph{Towards Meta-Learning to Study Plasticity}
Biological systems are intrinsically homeostatic and the study of such systems has been limited to manually designed studies.
The automatic homeostatic mechanism 
(\cref{eq:paramsovertime,eq:forget})
could, in principle, be fully implemented in a biological system, though we do not expect biology to explicitly implement gradient descent.
\par
The study of biological mechanism that underlie neural changes, ie plasticity, could thus be aided by gradient-based systems. As plasticity is the underlying mechanism for learning, finding mechanisms that perform activity dependent plasticity is effectively a form of meta-learning.
\par
To show how our method applies to this scenario we meta-learn a homeostatic plasticity mechanism, whereby a given target firing frequency $f_{tgt}$, dynamically adapts the input gain and channel densities to bring the neurons to behave in a specified regime.

\paragraph{
Emulating Unidentified Neural Mechanisms}
An alternative to the previously introduced meta-learning approach to homeostatic plasticity, is to study online automatic  homeostatic control directly as a black-box substitute for cellular plasticity.
One of the observed plasticity mechanisms in the brain is spike-timing dependent plasticity (STDP). It has been theorized that such mechanism most likely does not exist directly in the cell itself, but it the result of an underlying learning mechanism in the brain.
Online homeostatic control can function itself as a black-box, unknown learning mechanism in the brain, that aims to achieve a certain goal.
Then, by invoking a similar STDP-protocol, we can observe how such method would manifest as potential experimental observations .

We find that STDP can be observed as a side-effect, from both frequency tuning in HH cells (Figure~\ref{fig:stdp}A-B), as well as Ca$^{2+}$-concentration tuning in IO cells (Figure~\ref{fig:stdp}C-F).
In simulation, this happens at time-scales that are comparatively large to experimentally observed STDP signatures.
However, if we increase the forgetfulness parameter $\lambda$, required for stability, we return to more normal time-scales.
As such, gradient diffusion is also a method to study the relation between neural mechanisms underlying higher-level cellular goals and their manifestation as low-level plasticity levels.

\section*{Discussion}
In this article we introduced \emph{gradient diffusion}, a methodology that facilitates the calculation of parameter gradients for any existing, unmodified model-and-neurosimulator combination, thereby enabling support for homeostatic control. This approach allows for the efficient online and offline tuning of realistic neuron models and the implementation of homeostatic mechanisms in large networks, with the overarching goal of developing more robust, composable, and adaptable brain models that elucidate both the slow and fast adaptation dynamics of the brain.

The fact that an unmodified simulator could be \emph{tricked} into calculating gradients has broader implications across computational science beyond neuroscientific applications. Many scientific disciplines use specialized simulation software for field-specific models. Due to the local nature of derivatives, gradient models share many characteristics with their source model. Similar tricks to gradient diffusion could thus be applied in different fields, leading to easier-to-tune models across scientific fields. 

An interesting property of the gradient model, beyond computational science, is its interpretation as pure local dynamics and diffusive ion-species. Non-physical, non-local communication is not required, which means that gradient calculation is possible within biological constraints of cellular mechanisms. This means that full gradient calculation is \emph{in principle} feasible within biology. While cells are unlikely to keep track of an exact parameter-Jacobian, certain processes could reflect the computed gradients introduced.

A  research opportunity arising from the method is deriving the brain's slow dynamics and learning mechanisms under the constraints of cellular homeostasis. As shown, training many parameters over extended periods allows \emph{black-box} models, such as neural networks representing protein networks, to act as homeostatic-control agents within cells. Promising results have emerged in simplified neuron models~\cite{najarro2020meta}. Transferring the approach to realistic neuron models creates a data-driven possibility to discover biological mechanisms with slow dynamics.

Another research opportunity, as shown, is directly using the method as a substitute for plasticity.
As gradients are implemented using methods that do not violate physical principles, this could even serve as a biological basis of unknown plasticity processes.
A wide range of plasticity mechanisms has been observed in nature, with their underlying goal unknown.
Assuming goal-directed online-tuning, we show that STDP mechanisms could be replicated just as they are known in biology.

Limitations of the method relate to the fact that brain models extend beyond the cable equation: spike propagation with delays, stochastic models, reaction kinetics and ion dynamics, and Nernst potentials are not included in the method. These can be solved mathematically but would require modifications to the brain simulators, such as annotating spikes in-transit with gradient vectors. 
This omission does not directly affect homeostatic tuning: as shown, by handling some parameter-dependent processes as constant, networks can still be tuned, as is also the case in the biological brain.
More complex tasks, like training a computational function in a network of cells, would require modification of existing simulators or development of a new, fully differentiable brain simulator.

Our methodology adopts forward propagation through time (FPTT), commonly employed in artificial-neural-network tuning. As demonstrated, this enables in-simulation parameter tuning and homeostatic control. Without the constraints imposed by transient-neurosimulator mechanics --characteristic of many simplified brain models -- a range of alternatives, such as backward propagation through time, surrogate gradients, reverse-time adjoint-methods could be explored~\cite{neftci2019surrogate,kidger2022neural}. However, implementing these would necessitate developing a novel neurosimulator from scratch. In this context, we are optimistic about the potential of future differentiable simulators.

\section*{Methods}
In this section, we explain how the gradient model is derived from the cable equations and evaluate the approach. We start with a description of the cable equation as used in brain-simulation software, followed by a short description of the \emph{sensitivity equation}. We then combine these equations to form the gradient model and discuss the stability of homeostatic mechanisms built on top of these gradients and the necessity of a forgetting parameter $\lambda$. We conclude with the evaluation methods.

\begin{figure*}[t]
\include{nmodl}
\vspace{-3em}
\caption{
  Implementation of the gradient model in the common NMODL interface.
    Right-hand side derivatives are to be calculated by symbolic of automatic gradient derivation. Derivatives in square brackets are to be interpreted as regular state variables that represent the respective gradients.
    xd is a vector of ions.
    Some freedom of notation was allowed:
    NMODL does not support vectors or matrices. Instead, such objects should be read as written as single scalar values (i.e. $\mathbf s = s_1, s_2... s_N$)
}\label{fig:nmodl}
\end{figure*}

\paragraph{Brain Simulation}
Computational neuroscience seeks to understand and explain brain dynamics and function by simulation and analysis of the electrophysiological behavior of biophysically meaningful models, which allow establishing causal connections across scales.

The simplest possible equation to model a neuron at a biophysical level is the integration of current $i$ over its capacitive membrane. This \emph{transmembrane} current is usually the sum of individual ionic-channel currents. The current through each ionic channel is calculated from the cell's voltage, channel internal state $\mathbf s$ and system parameters $\theta$.

However, biological neurons, have an intricate spatial structure: Typically, a single root (soma), an axon and a branched dendritic tree. Modelling this accurately requires modelling the spatial structure of these neurons. In effect, beyond current influx and outflux over the capacitive ($C_m$) cell membrane, we also model the \emph{longitudal} current along the branches of the neural structure (\cref{fig:cablecelldef}A, blue). This is visible as a diffusion operator ($\Delta_x/R_l$) in the voltage equations.

In modelling, these spatial trees are discretized into \emph{segments}, leading to \emph{multicompartmental} neural models. Each segment is treated as a tapered cylinder. The longitudal diffusion term takes this into account by considering the change in surface area with respect to longitudal distance $\left(\frac{\partial S}{\partial x}\right)$. Trivially, $\left(\frac{\partial S}{\partial x}\right) = 2\pi R$ for a regular cylinder radius. Thus, we are left with the following \emph{cable equation}~\cite{hines1997neuron,paper:arbor2019}
\begin{align}
    \frac{d}{dt} V &= \frac{\Delta_x V}{C_m R_l \left(\frac{\partial S}{\partial x}\right)} -
        \frac{i\left(V, \mathbf s, \theta \right)}{C_m} \label{eq:arbv} \\
    \frac{d}{dt} {\mathbf s} &=  \mathbf f \left(V, \mathbf s, \theta \right)
\end{align}

\paragraph{Brain-Simulation Software}
Biophysically realistic neurons are simulated using dedicated brain simulation platforms, including NEURON~\cite{carnevale2006neuron}, Arbor~\cite{paper:arbor2019}, EDEN~\cite{panagiotou2022eden}. These contain the necessarily patchwork of mechanism to define a neural model. A full mathematical model describing these does not exist.
Of great importance to the modeler is the interface used to describe the model to the simulator.
NEURON and Arbor use the NMODL language to define mechanism dynamics, while EDEN is built around the NeuroML language.
In Arbor, NMODL is compiled to a C++ \emph{mechanism catalogue}, which then interfaces with a common Application Binary Interface (ABI).
The cable equation in \cref{eq:arbv} is taken from the Arbor brain simulator, but the same equation is solved by NEURON or EDEN.
On the programming side, these brain simulators allow the user to specify the functions $
i\left(V, \mathbf s, \theta \right)$ and  $\mathbf f \left(V, \mathbf s, \theta \right)$ using the NMODL or NeuroML languages, the rest is handled by the simulator.

\paragraph{Sensitivity Equation}
Assuming general dynamical system $\dot x = f(x, \theta)$, we have the time derivative of the state-parameter gradient as the sensitivity equation (~\cite{gronwall1919note}):

\begin{align}
    \frac{\mathrm{d}}{\mathrm{d}t}
    \varjac{x}{\theta}
    =
    \frac{\partial f(x, \theta)}{\partial \theta}
    +
    \frac{\partial f(x, \theta)}{\partial x}
    \varjac{x}{\theta}
    \label{eq:sens}
\end{align}

where we denote with $\left[\cdot\right]$ the state matrix of this new \emph{gradient system}, for which we can explicitly solve using numerical integration.

\paragraph{Gradients for Brain Models}
The most obvious solution to calculate gradient for neural models is by automatic gradient calculation throughout the entire simulator~\cite{jones2024efficient,deistler2024differentiable}.

\begin{align}
    \frac{\partial V}{\partial \theta} = \frac{\partial }{\partial \theta} ODESolve\left(
        \dot V,
        \dot {\mathbf s}
    \right)
\end{align}

This does not work in practice, as automatic gradient calculation is not possible in any currently existing full-featured brain simulator. Writing an differentiable ODESolve function to handle brain models, would constitute to writing a new brain simulator from scratch, which is a long and costly effort. Instead, we implement a gradient simulators by co-simulating the \emph{gradient model} of a neuron, in parallel to the neuron which is simulated by the simulator of choice. Our method leverages the development effort and optimization put into published complex simulations and enables computational neuroscientists to tackle new problems requiring on the fly gradient computation and optimization.

Applying the sensitivity equation \cref{eq:sens} to the cable equation \cref{eq:arbv}, we obtain the gradients for the cable cell. As such, we can now express a second \emph{gradient model}, as:
\begin{align}
    \frac{d}{dt} {\varjac{V}{\theta}} &=
    \frac{\Delta_x \varjac{V}{\theta}}{C_m R_l \left(\frac{\partial S}{\partial x}\right)}
        - \frac1{C_m} \frac{\partial i}{\partial \theta}
        \nonumber \\&\phantom{=}
        - \frac1{C_m} \frac{\partial i}{\partial V} \varjac{V}{\theta}
        - \frac1{C_m}\frac{\partial i}{\partial \mathbf s} \varjac{\mathbf s}{\theta} \\
    \frac{d}{dt}{\varjac{\mathbf s}{\theta}} &=
                    \frac{\partial \mathbf f}{\partial \theta} +
                    \frac{\partial \mathbf f}{\partial V} \varjac{V}{\theta} +
                    \frac{\partial \mathbf f}{\partial \mathbf s} \varjac{\mathbf s}{\theta}
                    \label{eq:grad}
\end{align}

Notably, the longitudinal voltage-diffusion term led to a \emph{gradient-diffusion} term in the gradient model. Diffusion of state variables is not directly expressible in NEURON or Arbor. Instead, we can use built-in diffusion of custom ions, for example in Arbor~\cite{nora_abi_akar_2023_8233847}:

\begin{equation}
    \frac{d}{dt} c = \frac{\beta \Delta_x}{\left(\frac{\partial S}{\partial x}\right)} c + i_c
\end{equation}
\noindent by setting the ionic diffusivity $\beta = \frac{1}{C_m R_l}$ (in practice, $\beta = 100/C_mR_l$ after unit conversion). In this way, we can calculate gradients for multicompartmental cells against the normal Arbor user-level API.

The resulting gradient system mirrors adaptation properties of the underlying biological system. For instance, in the case of neurons, we can express the gradient system as a ion-channel mechanism and a set of extra diffusive ions.
Their dynamics can we derived by (automatic) gradient calculation from the existing neuron model.
A template NMODL file implementing the gradient model is shown in \cref{fig:nmodl}.

\paragraph{Homeostatic-Control Stability}
Automatic homeostatic mechanism require a stable control process. Here we model how stability is maintained under continuos gradient-based updates.

Consider a passive neuron model with a target potential and leak, and the reversal potential $E_{leak}$ as parameter $\theta$ to be tuned under an integrated square root error $e$:

\newcommand{\E}[1]{ E_{\mathrm{#1}} }
\begin{align}
    i &= g_{\mathrm{leak}} \left( V - \E{leak} \right)
    &
    \frac{d}{dt} {\mathbf s}  &= \frac12 \left( V - \E{target} \right)^2
    \\
    \theta &= \E{leak}
    &
    e &= s_0
\end{align}
\noindent which leads to the following nonlinear system of PDEs for a single linear neuron with no varying radius,
where we have now included an extra gradient-forgetting parameter $\lambda$ from \cref{eq:forget} to prevent stability problems (as will be shown next):

\begin{align}
    \frac{d}{dt} V &=
         \frac{\Delta_x V}{2 \pi R C_m R_l} -
         \frac{g_{\mathrm{leak}} \left( V - \E{leak} \right)}{C_m}
    \\
    \frac{d}{dt}{\varjac{V}{\theta}} &=
        \frac{\Delta_x \varjac{V}{\theta}}{2  \pi R C_m R_l}
        + \frac{g_{\mathrm{leak}}}{C_m} \left(
        1 - \varjac{V}{\theta}
        \right)
    \\
    \frac{d}{dt} {\mathbf \theta} &=
        - \alpha \varjac{\mathbf s}{\theta}
    \\
    \frac{d}{dt}{\varjac{\mathbf s}{\theta}}' &=
        -\lambda \varjac{\mathbf s}{\theta}'
        +
        \left( V - \E{target} \right) \varjac{V}{\theta}
\end{align}

After the initial transient we obtain $\varjac{V}{\theta} = 1$. Without loss of generality, we can set $\E{target} = 0$. If we now apply a spatial sinusoidal perturbation $V \propto \sin\left(\omega x\right)$, we obtain the linear system:

\begin{equation}
    \frac{d}{dt}
    \begin{pmatrix}
        V \\
        \varjac{\mathbf s}{\theta}' \\
        \E{leak}
    \end{pmatrix}
    =
    \begin{pmatrix}
        \frac{-g_{\mathrm{leak}}}{C_m}
        -
        \frac{\omega^2}{2 \pi R C_m R_l}
        & 0 & \frac{g_{\mathrm{leak}}}{C_m} \\
        1 & -\lambda & 0 \\
        0 & -\alpha & 0
    \end{pmatrix}
    \begin{pmatrix}
        V \\
        \varjac{\mathbf s}{\theta}' \\
        \E{leak}
    \end{pmatrix}
\end{equation}
\noindent and resulting Routh–Hurwitz stability criterion\cite{Sandrock_tbcontrol_A_python}:

\begin{align}
\alpha <
\frac{
\begin{matrix}
\left(\pi C_{m}^{2} R R_{\mathrm{leak}} \lambda \omega^{2} + 2 g_{\mathrm{leak}} \lambda\right) \cdot \qquad\qquad\qquad \\
\qquad\qquad\qquad \left(\pi C_{m}^{2} R R_{l} \omega^{2} + 2 C_{m} \lambda + 2 g_{\mathrm{leak}}\right)
\end{matrix}
}{4 C_{m} g_{l}}
\end{align}

\noindent which has a lower bound as $\alpha < \lambda \left(\lambda +  \frac{g_{l}}{C_{m}}\right)$ in the case $\omega=0$ (or a single point neuron). In other words: to have stable online learning, the gradient needs to forget the components already reflected in parameters updates. Further spatial perturbations do not make the system more unstable. The same analysis can be repeated by adding a $\beta_{\theta} \Delta_x \theta$ diffusion term to the theta equation as in \cref{fig:nmodl}, where again spatial perturbations do not lead to a more unstable system.

\paragraph{Arbor-Pycat Library}
Crucial in the method is availability of gradients for the current and ode system, greatly amplified by automatic gradients calculation systems - eg. JAX. JAX, as an ML-library, fully expresses the desired interface for expressing internal dynamics of neural models~\cite{landsmeer2024tricking}. To ease implementation, a python wrapper around the C mechanism ABI was written that exposed the mechanism internals as numpy-style buffers, allowing JAX to work with it. This is released as a installable python package \emph{Arbor-Pycat}~\cite{llandsmeer2024pycat}.

\paragraph{Axon model}
The Hodgkin-Huxley model was implemented in Arbor-Pycat, as well as its gradient system using \cref{eq:grad}.  An axon was defined of 11 compartments, using the resulting mechanism. A current pulse was applied at t=200ms and multiplied by the parameter $\theta=w$ before being injected into one end of the axon.

\paragraph{Inferior Olivary Nucleus model}
A modified version~\cite{negrello2019quasiperiodic} of the de Gruijl model~\cite{de2012climbing} was implemented in Arbor-Pycat. The morphology of mouse IO neuron C4A was used~\cite{vrieler2019variability}. After discretization, it contains 4 axonal, 15 dendritic and 2 somatic compartments. Initial parameters are listed in Table~\ref{tab:initparamio}. For programming ease, these default values were scaled by the parameter $\theta$, which was thus a vector of ones, initially. Optimization was performed against the loss-function:

\begin{align}
    \mathcal{L} = \frac{1}{T}\int_{0}^T\left(v(t) - \hat v(t)\right)^2dt
\end{align}

\noindent from which updates to the state were calculated from \cref{eq:gupdate}. The Adam optimizer \cite{kingma2014adam} with learning rate $10^{-2}$ was used, via the optax library \cite{deepmind2020jax}.  This was compared to the the CMA-ES\cite{hansen2019pycma} algorithm with population size 20 and initial $\sigma=0.1$. Initial parameters are listed in \cref{tab:initparamio}.

\paragraph{Homeostatic control of spike frequency}
Experiments on the usability of gradients for homeostatic control were performed in the Hodgkin-Huxley model. The target was a given spiking frequency, while a Ornstein-Uhlenbeck noise input was given as current input. Tunable parameters were the conductance values of the sodium and potassium channels, and the input scaling. 
To also show the usability of the method through standard NMODL interface, the Hodgkin-Huxley model was implemented symbolically in SymPy. The SymPy library was then used to implement the gradient model in NMODL code.

The online frequency-tuning task requires a differentiable measure of spike frequency that is calculable from the cell state. For the Hodgkin-Huxley system, this cell state are the gating variables $\mathbf s = \left(m, h, n\right)$. We hypothesized that a robust metric could be the amount of revolutions in the gating-variable phase plane. Preliminary visual inspection suggested $(m,\dot m)$ to be one of the more optimal choices, but the extra time dependence $\dot m$ would add a lot of compute. Instead, it was decided to count the amount of revolutions in the $(m,h)$ phase plane:

\begin{equation}
    N = \frac{1}{2\pi}
    \operatorname{unwrap}\left(\operatorname{arctan2}\left(m-\frac12, h-\frac12\right)\right)
\end{equation}

\noindent which derivative (\cref{eq:dotN}) is low-passed (\cref{eq:dotNlp}) to obtain a differentiable frequency metric.
The final instantaneous error-function is then the squared difference to the target frequency (\cref{eq:ferror}):

\begin{align}
    \frac{d}{dt} N &= \frac{1}{2\pi}
             \frac{\left(h-\frac12\right)\frac{d}{dt} m - \left(m - \frac12\right)\frac{d}{dt} h}
                  {\left(h-\frac12\right)^2 + \left(m - \frac12\right)^2}
                  \label{eq:dotN}
    \\
    \tau_f \frac{d}{dt} f &= \frac{d}{dt} N - f
    \label{eq:dotNlp}
    \\
    \frac{d}{dt} e &= \frac12\left(f - f_{\mathrm{tgt}}\right)^2
    \label{eq:ferror}
\end{align}

\noindent Here, $\tau_f$ is preferable on a time scale much larger than the expected spiking frequency (e.g. $\tau_f \approx 10/f_{\mathrm{tgt}}$)

\paragraph{Hardware}
AMD Ryzen 9 3900X 12-Core processor and NVIDIA RTX 6000 GPU

\newcommand{\T}[3]{{\makebox[2.5em][l]{${#1}_{\mathrm{#2}}$}$=#3$}}
\begin{table}
    \centering
    \begin{tabularx}{\columnwidth}{llll}
         \toprule
         Soma (S/cm$^2$) & Dendrite (S/cm$^2$)   \\
         \midrule
    \T{g}{leak}{1.3\times 10^{-5}} & \T{g}{leak}{1.3\times 10^{-5}} \\
    \T{g}{CaL}{0.045} & \T{g}{KCa}{0.220}\\
    \T{g}{Na}{0.030} & \T{g}{CaH}{0.010}\\
    \T{g}{Kdr}{0.030} & \T{g}{h}{0.015}\\
    \T{g}{K}{0.015} &                   \\
         \midrule
        Axon (S/cm$^2$) & Potential (mV) \\
         \midrule
         \T{g}{leak}{1.3\times 10^{-5}} & \T{V}{leak}{\phantom{-0}10} \\
         \T{g}{Na}{0.200}     & \T{V}{Ca}{\phantom{-}120} \\
         \T{g}{K}{0.200}     & \T{V}{Na}{\phantom{-0}55} \\
                                  & \T{V}{K}{-\phantom{0}75} \\
                                  & \T{V}{h}{-\phantom{0}43} \\
         \bottomrule
    \end{tabularx}
    \caption{Initial parameters for inferior olivary cell model}\label{tab:initparamio}
\end{table}

\paragraph{Acknowledgments}
This paper is partially supported by the European-Union Horizon Europe R\&I program through projects SEPTON (no. 101094901) and SECURED (no. 101095717), the NWO - Gravitation Programme DBI2 (no. 024.005.022) and through the Erasmus MC Convergence Health and Technology Integrative Neuromedicine Flagship Program.. The RTX6000 used for this research was donated by the NVIDIA Corporation. 

\bibliography{main}

@INPROCEEDINGS{
    paper:arbor2019,
    author={N. {Abi Akar} and B. {Cumming} and V. {Karakasis} and A. {Küsters} and W. {Klijn} and A. {Peyser} and S. {Yates}},
    booktitle={2019 27th Euromicro International Conference on Parallel, Distributed and Network-Based Processing (PDP)},
    title={{Arbor --- A Morphologically-Detailed Neural Network Simulation Library for Contemporary High-Performance Computing Architectures}},
    year={2019}, month={feb}, volume={}, number={},
    pages={274--282},
    doi={10.1109/EMPDP.2019.8671560},
    ISSN={2377-5750}}

@misc{nora_abi_akar_2023_8233847,
  author       = {Nora Abi Akar and
                  John Biddiscombe and
                  Benjamin Cumming and
                  Marko Kabic and
                  Vasileios Karakasis and
                  Wouter Klijn and
                  Anne Küsters and
                  Alexander Peyser and
                  Stuart Yates and
                  Thorsten Hater and
                  Brent Huisman and
                  Espen Hagen and
                  Robin De Schepper and
                  Charl Linssen and
                  Harmen Stoppels and
                  Sebastian Schmitt and
                  Felix Huber and
                  Max Engelen and
                  Fabian Bösch and
                  Jannik Luboeinski and
                  Simon Frasch and
                  Lukas Drescher and
                  Lennart Landsmeer},
  title        = {Arbor Library v0.9.0},
  month        = nov,
  year         = 2023,
  publisher    = {Zenodo},
  version      = {v0.9.0},
  doi          = {10.5281/zenodo.8233847},
  url          = {https://doi.org/10.5281/zenodo.8233847}
}

@article{landsmeer2024tricking,
  title={Tricking AI chips into simulating the human brain: A detailed performance analysis},
  author={Landsmeer, Lennart PL and Engelen, Max CW and Miedema, Rene and Strydis, Christos},
  journal={Neurocomputing},
  pages={127953},
  year={2024},
  publisher={Elsevier}
}

@article{jones2024efficient,
  title={Efficient optimization of ODE neuron models using gradient descent},
  author={Jones, Ilenna Simone and Kording, Konrad Paul},
  journal={arXiv preprint arXiv:2407.04025},
  year={2024}
}

@article{de2012climbing,
  title={Climbing fiber burst size and olivary sub-threshold oscillations in a network setting},
  author={De Gruijl, Jornt R and Bazzigaluppi, Paolo and de Jeu, Marcel TG and De Zeeuw, Chris I},
  journal={PLoS computational biology},
  volume={8},
  number={12},
  pages={e1002814},
  year={2012},
  publisher={Public Library of Science San Francisco, USA}
}

@article{negrello2019quasiperiodic,
  title={Quasiperiodic rhythms of the inferior olive},
  author={Negrello, Mario and Warnaar, Pascal and Romano, Vincenzo and Owens, Cullen B and Lindeman, Sander and Iavarone, Elisabetta and Spanke, Jochen K and Bosman, Laurens WJ and De Zeeuw, Chris I},
  journal={PLoS computational biology},
  volume={15},
  number={5},
  pages={e1006475},
  year={2019},
  publisher={Public Library of Science San Francisco, CA USA}
}

@article{vrieler2019variability,
  title={Variability and directionality of inferior olive neuron dendrites revealed by detailed 3D characterization of an extensive morphological library},
  author={Vrieler, Nora and Loyola, Sebastian and Yarden-Rabinowitz, Yasmin and Hoogendorp, Jesse and Medvedev, Nikolay and Hoogland, Tycho M and De Zeeuw, Chris I and De Schutter, Erik and Yarom, Yosef and Negrello, Mario and others},
  journal={Brain Structure and Function},
  volume={224},
  number={4},
  pages={1677--1695},
  year={2019},
  publisher={Springer}
}

@misc{hansen2019pycma,
  author       = {Nikolaus Hansen and Youhei Akimoto and Petr Baudis},
  title        = {{CMA-ES/pycma} on {G}ithub},
  howpublished = {Zenodo, DOI:10.5281/zenodo.2559634},
  month        = feb,
  year         = 2019,
  doi          = {10.5281/zenodo.2559634},
  url          = {https://doi.org/10.5281/zenodo.2559634},
}

@article{hines1997neuron,
  title={The NEURON simulation environment},
  author={Hines, Michael L and Carnevale, Nicholas T},
  journal={Neural computation},
  volume={9},
  number={6},
  pages={1179--1209},
  year={1997},
  publisher={MIT Press}
}

@book{carnevale2006neuron,
  title={The NEURON book},
  author={Carnevale, Nicholas T and Hines, Michael L},
  year={2006},
  publisher={Cambridge University Press}
}

@article{panagiotou2022eden,
  title={EDEN: A high-performance, general-purpose, NeuroML-based neural simulator},
  author={Panagiotou, Sotirios and Sidiropoulos, Harry and Soudris, Dimitrios and Negrello, Mario and Strydis, Christos},
  journal={Frontiers in neuroinformatics},
  volume={16},
  pages={724336},
  year={2022},
  publisher={Frontiers Media SA}
}

@article{gronwall1919note,
  title={Note on the derivatives with respect to a parameter of the solutions of a system of differential equations},
  author={Gronwall, Thomas Hakon},
  journal={Annals of Mathematics},
  volume={20},
  number={4},
  pages={292--296},
  year={1919},
  publisher={JSTOR}
}

@misc{llandsmeer2024pycat,
  author    = {Lennart Landsmeer},
  title     = {arbor-pycat},
  version   = {1.0},
  publisher = {Zenodo},
  month     = Dec,
  year      = 2024,
  doi       = {10.5281/zenodo.14506753},
  url       = {http://dx.doi.org/10.5281/zenodo.14506753}
}

@article{najarro2020meta,
  title={Meta-learning through hebbian plasticity in random networks},
  author={Najarro, Elias and Risi, Sebastian},
  journal={Advances in Neural Information Processing Systems},
  volume={33},
  pages={20719--20731},
  year={2020}
}

@software{Sandrock_tbcontrol_A_python,author = {Sandrock, Carl},license = {GPL-3.0},title = {{tbcontrol: A python library for textbook control problems}},url = {https://github.com/alchemyst/Dynamics-and-Control}}

@software{deepmind2020jax,
  title = {The {D}eep{M}ind {JAX} {E}cosystem},
  author = {DeepMind and Babuschkin, Igor and Baumli, Kate and Bell, Alison and Bhupatiraju, Surya and Bruce, Jake and Buchlovsky, Peter and Budden, David and Cai, Trevor and Clark, Aidan and Danihelka, Ivo and Dedieu, Antoine and Fantacci, Claudio and Godwin, Jonathan and Jones, Chris and Hemsley, Ross and Hennigan, Tom and Hessel, Matteo and Hou, Shaobo and Kapturowski, Steven and Keck, Thomas and Kemaev, Iurii and King, Michael and Kunesch, Markus and Martens, Lena and Merzic, Hamza and Mikulik, Vladimir and Norman, Tamara and Papamakarios, George and Quan, John and Ring, Roman and Ruiz, Francisco and Sanchez, Alvaro and Sartran, Laurent and Schneider, Rosalia and Sezener, Eren and Spencer, Stephen and Srinivasan, Srivatsan and Stanojevi\'{c}, Milo\v{s} and Stokowiec, Wojciech and Wang, Luyu and Zhou, Guangyao and Viola, Fabio},
  url = {http://github.com/google-deepmind},
  year = {2020},
}

@article{kingma2014adam,
  title={Adam: A method for stochastic optimization},
  author={Kingma, Diederik P},
  journal={arXiv preprint arXiv:1412.6980},
  year={2014}
}

@article{deistler2024differentiable,
  doi = {10.1101/2024.08.21.608979},
  year = {2024},
  publisher = {Cold Spring Harbor Laboratory},
  author = {Deistler, Michael and Kadhim, Kyra L. and Pals, Matthijs and Beck, Jonas and Huang, Ziwei and Gloeckler, Manuel and Lappalainen, Janne K. and Schr{\"o}der, Cornelius and Berens, Philipp and Gon{\c c}alves, Pedro J. and Macke, Jakob H.},
  title = {Differentiable simulation enables large-scale training of detailed biophysical models of neural dynamics},
  journal = {bioRxiv}
}

@article{almog2016realistic,
  title={Is realistic neuronal modeling realistic?},
  author={Almog, Mara and Korngreen, Alon},
  journal={Journal of neurophysiology},
  volume={116},
  number={5},
  pages={2180--2209},
  year={2016},
  publisher={American Physiological Society Bethesda, MD}
}

@article{neftci2019surrogate,
  title={Surrogate gradient learning in spiking neural networks: Bringing the power of gradient-based optimization to spiking neural networks},
  author={Neftci, Emre O and Mostafa, Hesham and Zenke, Friedemann},
  journal={IEEE Signal Processing Magazine},
  volume={36},
  number={6},
  pages={51--63},
  year={2019},
  publisher={IEEE}
}

@article{kidger2022neural,
  title={On neural differential equations},
  author={Kidger, Patrick},
  journal={arXiv preprint arXiv:2202.02435},
  year={2022}
}

\end{document}